\documentclass[pre,aps,twocolumn,floats]{revtex4} 
\input epsf
\usepackage{bm}
\usepackage{multirow}
\usepackage{amsmath}
\usepackage{graphicx}
\usepackage{dcolumn}
\usepackage{bm,color,soul,ulem}
\usepackage{url,hyperref,varioref}
\hypersetup{colorlinks,citecolor=blue,linkcolor=red,urlcolor=blue}

\begin{document}

\title{Membrane budding driven by intra-cellular ESCRT-III filaments}
\author{Sk Ashif Akram}
\author{Gaurav Kumar}
\author{Anirban Sain}
\email{asain@phy.iitb.ac.in}
\affiliation{ Physics Department, Indian Institute of  Technology-Bombay, 
Powai, Mumbai, 400076, India. }

\begin{abstract}
Exocytosis is a common transport mechanism via which cells 
transport out non-essential macromolecules (cargo) into the extra cellular space. 
ESCRT-III proteins are known to help in this. They polymerize into a conical spring like structure and help deform the cell membrane 
locally into a bud which wrapps the outgoing cargo. 
we model this process using a continuum energy functional. It consists of
elastic energies of the membrane and the semi-rigid ESCRT-III filament, 
favorable adhesion energy between the cargo and the membrane, and affinity 
among the ESCRT-III filaments. We take the free energy minimization route to identify the sequence of composite structures which form 
during the process. We show that membrane adhesion of the cargo is the driving force for this budding process and not the buckling of ESCRT-III filaments 
from flat spiral to conical spring shape. 
However ESCRT-III stabilizes the bud once it forms. Further we conclude that a nonequilibrium process is needed to pinch off/separate the stable bud (containing the cargo) from the cell body. 
\end{abstract}
\maketitle

\begin{figure*}
\centering
\includegraphics[width=.9\linewidth, angle=0]{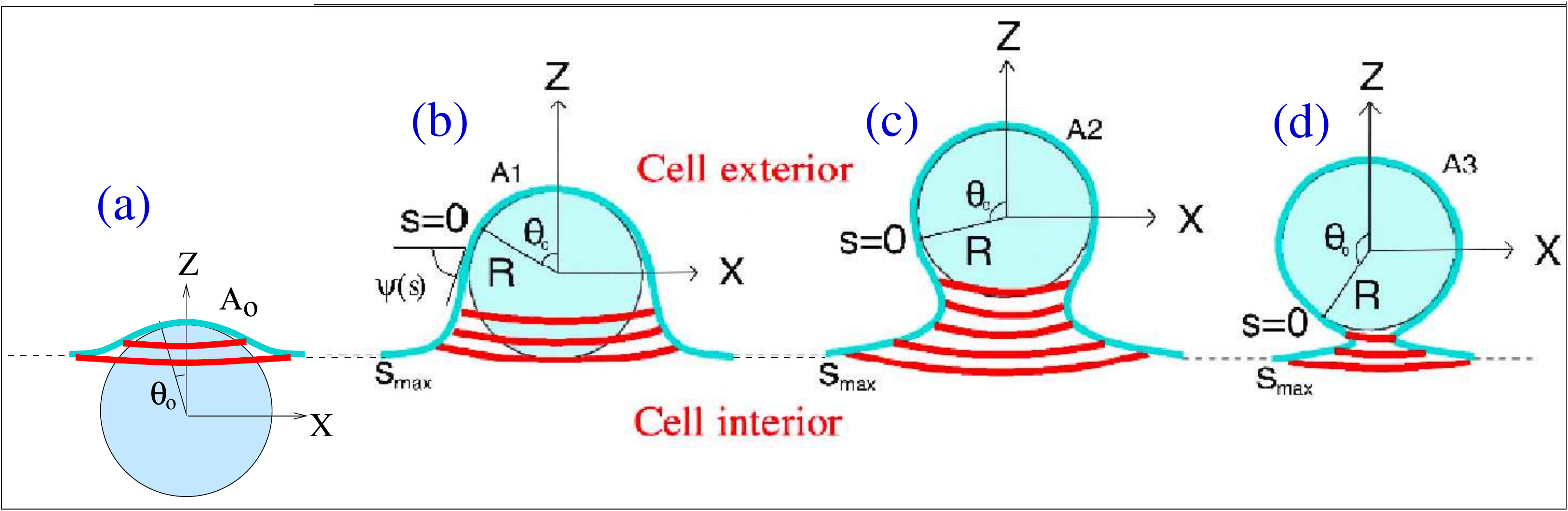}
\caption { Schematic diagram showing progression of ESCRT-III mediated exocytosis in our model. The spherical cargo is 
partially wrapped by membrane (in blue) and unwrapped part of 
the membrane bud is covered by ESCRT-III filaments (in red). Degree of wrapping of the cargo is parameterised by the angle $\theta_0$
and the area of the bud gradually increases from left to right ($A_0\rightarrow A_3$). Unwrapped portion of the membrane is 
parameterised by the contour $s$, when $s\in [0,s_{max}]$. }
\label{fig.schematic}
\end{figure*} 

\section{Introduction} 
Eukaryotic cells use various proteins, which deform the cell membrane, in order to allow transport of essential  macromolecules (cargo) between the cell and its exterior. ESCRT-III is one such protein in the 
protein family {\it Endosomal sorting complex required for transport} or ESCRT.  It  helps in the formation of a membrane bud and its detachment, carrying non-essential cargo out of the cell. One of the unique abilities 
of ESCRT-III filaments compared to other intracellular filaments is 
its ability to severe the membrane neck from inside the membrane bud. 
Dynamin filaments, for example wraps around the neck from outside in 
order to severe it. 
ESCRT-III are intracellular filamentous proteins which bind to the 
cell membrane from inside, around the cargo in the form of flat spirals.
In the absence of any direct visualisation of the process in vivo, various proposals have emerged about the dynamical process.
It is not clear whether the flat spiral grows after the cargo attaches to
the already deformed membrane or whether the flat ESCRT-III spiral grows first and the cargo expels the spiral and attaches itself to the membrane.
Irrespective of the initiation of the spiral, it buckles into a helical shape and push the cargo out of the cell. Subsequently the spiral also retracts itself back to the cell and disintegrates. During this process the cargo gradually wraps itself around with the cell membrane in the form of a  membrane bud and the bud eventually pinches off. It is not clear whether the process is driven by the energy gain due to the shape transformation of the ESCRT-III coil \cite{lenz2009membrane} 
or due to the adhesion of the cargo molecule to the cell membrane \cite{Lipowsky2016stabilization, Lipowski2018domes}.


 The cargo, which is assumed to be a sphere here, adheres to the cell membrane from the cell-interior and deforms it, see Fig.\ref{fig.schematic} The angular extent of adhering membrane area is marked by the polar angle $\theta_0$. The amount of membrane area adhering with the cargo is then $A = 2 \pi R^2 (1 - \cos \theta_0)$. The ESCRT-III filament grows around the cargo, initially as a flat spiral (not shown here). As the cargo wraps itself with more membrane area (i.e., increasing $\theta_0$), the filament lifts off as a helical spiral, shown in (a and b). As $\theta_0$ crosses $\pi/2$ the spiral is pressed in the middle as a neck develops, shown in (c). As $\theta_0$ approaches $\pi$ the conical spiral begins to retract from the neck and moves towards a pinch-off configuration. From (a) through (c) the adhering area  grows as $A_0\rightarrow A_1 \rightarrow A_2\rightarrow A_3$. The part of the surface not adhering to the sphere is parameterised \cite{PhysRevA.44.1182,julicher1994shape,PhysRevE.94.062404} by the axisymmetric curve $\vec r(s)=[X(s),Z(s)]$ which extends from $s=0$ to $s=s_{max}$. The location of the point where the sphere looses contact with the membrane is $X(s=0) = R \sin \theta_0$ and $Z(s=0) = Z_0 + R \cos \theta_0$, where $(X_0,Z_0)$ is the center of the sphere. The shape of this surface is expressed using an angle $\psi(s)$,  which is the angle between the local tangent to the curve and the $X$ axis, see fig(a). Note that $\psi(s=0) = \theta_0$ and we assume that $\psi (s_{max})=0$, i.e., it joins smoothly with the cell surface which is assumed to be flat. This is justified because the radius of the cell is much larger than the radius of the bud $R$.

It is also not clear if the attachment of flat spirals precedes attachment of cargo to the membrane. Question arises if the membrane is already covered with dense arrays of ESCRT-III flat spirals, then how does the cargo attach onto the membrane. On the other hand, if the cargo attaches to the membrane first, the ESCRT-III spirals will be excluded from that patch of the membrane. However it is plausible that ESCRT-III can still polymerize around the patch forming bigger circles. 

Lenz et al had shown \cite{lenz2009membrane} that due to its intrinsic curvature and its favourable association with the membrane, the ESCRT-III filaments can undergo a buckling transition from a flat spiral to a helical spiral, all the while keeping its contact with the membrane. They assumed an attractive interaction between the turns of the ESCRT-III spral and the membrane of strength $-\mu$ per unit area. In their model the tightly packed helical shaped membrane was approximated as a stack of circular layers since the pitch is much smaller than the radius of the 
helix. We essentially use a variant of their model  to study the ESCRT-III mediated budding and exocytosis, which in addition to the filament instability also requires separation of the bud from the cell. In this context ref\cite{Lipowski2018domes} has highlighted the role of the adhesion energy between the cargo and the cell membrane in  the budding process. In our model below we include this adhesion energy contribution which turns out to be the main driver of exocytosis. In Fig.\ref{fig.schematic} we show a schematic diagram for the process.

Concerning the  transition of the flat ESCRT-III spiral into a helical spiral, a recent study has proposed a conformational change in the binding between the ESCRT-III molecules and the membrane \cite{ Saric2019changes, saric2020}.  They showed that this effect causes successive helical turns in the spiral to have decreasing radii, thereby  inducing the flat spiral to become a conical helix. However there is no direct experimental evidence for this conformational transition.   

The shape of the membrane bud during its development is not clear either.
The portion of the membrane joining the neck to the basal surface of the
cell has been argued to be cone shaped \cite{Lipowski2018domes} rather 
than dome shaped \cite{kozlov2009}. Ref\cite{Lipowski2018domes}, which 
has focused on the bud scission rather than development of the bud, has
shown that a cone shaped neck requires lesser adhesion energy, between
the filament and membrane, for scission compared 
to if it is dome shaped. Further, the process of fission of
the neck has been proposed to be a non-equilibrium active process
\cite{Saric2019changes} which serves to cross a free-energy barrier. 

Ref \cite{Lenz2015relaxation}, motivated by in vitro experiments on membrane covered with an array of flat ESCRT-III  coils (spirals),  have proposed a compressed coil hypothesis. In this in vitro experiment array the coils seem to be in a compressed state. This is inferred from the observation that often the outer layers of the neighboring coils meet at straight lines exhibiting an approximate hexagonal arrangement of coils. At lower concentration the coils are circles touching each other. Ref \cite{Lenz2015relaxation} argues that 
beyond a threshold of compression level the coils buckle out of the plane. 
However, it is not evident whether the flat spirals are in a compressed state 
in the in vivo situation \cite{hansenJBC08}. 

As mentioned earlier, we work with an augmented version of the model by Lenz et al. \cite{lenz2009membrane}. We focus on the development of the bud rather than its scission. We minimize the equilibrium free energy to look for the optimum state where the bending energy of the deformed membrane,  
adhesion energy of the cargo and the bending energy of the filament are jointly minimized. Description of the model is given below:

\section{Model}
As shown in figure \ref{fig.schematic} 
a spherical shaped outgoing cargo adheres to the dome shaped membrane at the top \cite{Lipowski2018domes,   fabrikant2009computational,  julicher1994shape, Lenz2015relaxation,  Lipowsky2016stabilization, PhysRevA.44.1182, PhysRevE.94.062404,rozycki2012membrane, Sabyasachi2013wrapping,Saric2019changes}. Our goal is to find an optimized surface configuration of the membrane and its evolution during buckling.  Dense spiral filaments (attached to the membrane) are simplified to be concentric circular rings with uniform surface density throughout the evolution. The equilibrium shape of the membrane, with its ESCRT filaments and the adhering cargo attached to it, is found by minimizing the free energy of the composite system. To describe the composite system consisting of a membrane and polymer, we use a model similar to that in \cite{lenz2009membrane}. We study the process using an equilibrium membrane elasticity based continuum model \cite{lenz2009membrane} (below) which includes the elasticity of the ESCTR-III filaments, lateral attraction between the filaments, adhesion between the filament and the membrane 
and adhesion between the membrane and the cargo.The bending energy cost of the membrane and the ESCRT filaments are given by the Helfrich energy and worm-like chain model, respectively. The adhesion energy of the cargo is an addition to the model.


We take into account the membrane-filament attraction, but we do not go into the details of the molecular structure of filaments since we are only interested in their overall coarse-grained contribution to the total energy of the system. A spherical bud gets created and eventually leads to the scission from larger membrane at the neck (the minimum radius region). The adhesion energy between the cargo and the membrane works to overcome the potential barrier induced by rest of the terms, thus making buckling process energetically favourable. The expression of the free energy is given by, $E_T = E + E_c$. Here $E_c$ is proportional to the
cap area of the membrane bud which partially wraps the cargo and it consists of favorable adhesion energy between the cargo and the membrane, and the elastic bending energy cost of the membrane. $E$ corresponds to the rest of the bud, below the cargo, which is covered by ESCRT-III filaments.  
\begin{equation}
     E_c= - 2 \pi  R^2 (\sigma _0 -\frac{\kappa}{2R^2}) (1-\cos \theta _0)
     \label{eq.Ec}
     \end{equation}
     and,
     \begin{eqnarray}
     E&=&\int_0^{s_{\max }} ds\Bigg{[}2 \pi  X (s)\Bigg{\{}\frac{k}{2} \left(\frac{1}{X(s)}-\frac{1}{\chi _0}\right)^2 \nonumber\\
     &+&\frac{\kappa}{2}\left(\frac{\sin \psi (s)}{X(s)}+\psi '(s)\right)^2  +(\sigma_1 -\mu )\Bigg{\}}\nonumber \\ 
     &+&\lambda (s) \Big{\{}X '(s)-\cos \psi (s)\Big{\}} + \eta(s)\Big{\{}Z '(s)+\sin \psi (s)\Big{\}}\, \Bigg{]} \nonumber \\
     &+& 2 \pi  \gamma [X(0) + X(s_{max})] 
\label{eq.E}
     \end{eqnarray}

Here, $k$ is bending stiffness of the filament, $\kappa$ the bending modulus of the membrane, $-\mu$ the attractive chemical potential between the filaments and the membrane, $\sigma_1$ is the surface tension and $\sigma_0$ is the attractive adhesion coefficient between the membrane and the spherical shaped cargo.

The first term represents the bending energy of the protein filament 
ESCRT-III. The tightly wound spiral is approximated as a stack of 
circular coils of finite thickness with variable radii $X(s)$,
while  $\chi_0$ is the intrinsic radius of curvature of these coils. The radius and the height variables $X(s)$ and $Z(s)$ represent the 
space curve  ($X(s),Z(s)$) parameterized by the contour length $s$, running between $0$ to $s_{max}$. An azimuthally symmetric deformed 
membrane (the partial bud) is the surface of revolution of this space 
curve  ($X(s),Z(s)$).  
One important assumption implicit here is that the filaments are closely packed and dense enough to cover whole of the surface. We assume that the filament density per unit area of the membrane remains uniform during the process. Note that this term will favor a helix of radius $\chi_0$ as opposed to a flat coil where all the concentric layers, except the one 
with radius $\chi_0$ will be strained and therefore have an energy cost.     

The second term denotes the Helfrich free energy of the deformed membrane represented by the space curve $\vec r(s)=[X(s),Z(s)]$.   
$\psi(s)$ is the angle between the tangent and the $X$ axis. All derivatives here are with respect to curve length $s$. The two $\psi$ dependent terms are the local principal curvatures on the azimuthally symmetric surface
Any sharp bending of the membrane, for example at the neck of the bud, costs high bending energy.  

The third term, proportional to the membrane area which is in contact with the filament, includes effect of surface tension $\sigma_1$ of the membrane and the attractive interaction $-\mu$ between the filament and the membrane, also known as membrane affinity of the filament. 
Since $\sigma_1-\mu$ is what matters we define $\sigma_1-\mu=-\mu'$ and look for variation of
shapes with respect to $\mu'$. Note that higher $\mu'$ will favor larger membrane area covered with the filament which requires longer filament
 length as well. This will promote the membrane bud to have a wider base
 rather than lifting off higher in the $Z$ direction. 
Lagrange multipliers $\lambda(s)$ and $\eta(s)$ ensure the relations between $X(s),Z(s)$ and $\psi$ \cite{PhysRevA.44.1182}. 
The next term is crucial to include the attractive interactions between the filaments (which are closely packed) and characterized by the line tension $\gamma$. 

The second last term involving $\gamma$ captures the lateral attraction between
the filaments \cite{lenz2009membrane} which show up as line tensions at the
two open edges of the helical coil. These open edges are essentially 1-D interface between membrane covered with filaments and the bare membrane.
Just as 2D surface tension at a liquid-air interface emerge from the fact
that liquid molecules have greater attraction among themselves (cohesion) 
than that with the air molecules, here the attraction between the ESCRT-III
filaments give rise to line tension which is positive definite like surface
tension. This term can be rewritten as 
\begin{equation}
2 \pi\gamma [X(0)+X(s_{max})]=2\pi\gamma\Big[\int_0^{s_{\max}}\hspace{-.5cm}X '(s) ds+ 2X(0)\Big]
\end{equation}
Note that for a given $\theta_0$ the term $X(s_{max})$ is dependent on the 
shape which is unknown apriori, but $X(0)$ is fixed. 

The last term combines the adhesion interaction energy between the surface of the cargo and the membrane wrapping around it, and bending energy of this part of the membrane with curvature $1/R$. Both the terms are proportional to the area of the partial sphere (dependent on $\psi(0)=\theta_0$), which is covered by the membrane. Both $\sigma_0$ 
and $\kappa/R^2$ has the unit of energy per unit area. The adhesion energy is negative (since attractive) while the bending energy is positive. 
We define an effective adhesion energy $\sigma= \sigma_0-\frac{\kappa}{2r^2}$ and assume $\sigma >0$. Note that this adheson energy term monotonically decreases as $\theta_0$ increases and hence promotes complete wrapping of the cargo by the membrane. 

As Fig.\ref{fig.schematic} shows, the evolving bud can be parameterized by
the angle $\theta_0$. Therefore $\theta_0$ serves as a proxy for time in 
our investigation. For different $\theta_0$ values, we solve for the lowest free energy configuration of the system. The path connecting these different configurations represent the quasi-static evolution of the system in the configuration space. We track how the curve length $s_{max}$ and the area of the unwrapped portion of the bud $A$ in the minimum energy configuration change as $\theta_0$  is varied. Thereby we generate the
potential energy profiles for the system as a function of $\theta_0$,
for a given set of parameters. 

To simplify calculation, we use dimensionless parameters and variables.  We use membrane  bending modulus $\kappa$ and the fixed radius of the spherical cargo $R$, as the characteristic energy and length scales, respectively, to make everything (including curve length $s$) dimensionless. In units of $\kappa$  ($=20 k_B T$) the relevant energy density $\mathcal E$ is now given by,

\begin{eqnarray}
    \mathcal{E}&=& 2 \pi  X(s) \Big[\frac{k}{2} \Big(\frac{1}{X(s)}-\frac{1}{\chi_0}\Big)^2 +\frac{1}{2} \Big(\frac{\sin \psi (s)}{X (s)}+\psi '(s)\Big)^2 \nonumber\\
    &-& \mu' \Big] + 2\pi\gamma X'(s)
    + \lambda (s)[X '(s)-\cos \psi (s)]
\end{eqnarray}

Here we retained the same notations for the all the non-dimensionalised quantities.
The corresponding Euler-Lagrange equations are:
\begin{eqnarray}
&&-\frac{k}{\chi_0^2}+2 \mu' -\psi '^2 + \frac{\left(k+\sin ^2\psi \right)}{X^2}+
    \frac{\lambda '(s)}{\pi}=0.\\
&&2 X ' \psi '+2 X \psi ''-\frac{\sin 2 \psi}{X}-\frac{\lambda (s)}{\pi} \sin \psi =0.\\
&& \cos \psi (s)-\chi '(s)=0.
\label{eq.EL}
\end{eqnarray}

The EL-equation that involves $Z$ is the constraint $Z'=-\sin \psi $ which is coupled with the above three equations via $\psi(s)$. But $\psi(s)$ can be obtained by solving the above three equations only. The constraint equation $Z'=-\sin \psi $ can be integrated later, with appropriate boundary conditions, to compute $Z(s)$, once $\psi(s)$ is known. 

The boundary conditions on the membrane shape are set by demanding continuity of slope between the partial sphere part  (adhering to the cargo) and the
part below it (attached to the filament), i.e., $\psi (0)= \theta_0$. Also the base of the bud is assumed to smoothly join onto the flat basal membrane,
i.e., $\psi (s_{max})=0$. The radius at the junction $\chi(0)$, where the membrane loose contact with the cargo is computed as $\chi(0)=\frac{\sqrt{4 \pi  A-A^2}}{2 \pi }$, where the wrapped area $A=2\pi (1 -\cos \theta_0)$ in units of $R=1$. Further, we assume $\lambda (0)=0$.  After computing $\psi(s)$ we compute the space curve by integrating the equations $Z'(s)=-\sin \psi (s)$ and $\chi'(s)=\cos \psi (s)$, where we use $Z(s_{max})=0$. Note that for a given $\theta_0$ we do not know $s_{max}$ a priori. Instead of extremizing the energy density with respect to $s_{max}$, which would increase number of 
equations to be solved by one we compute the lowest energy configurations
for different value of $s_{max}$ for a given $\theta_0$ and choose the
lowest energy configuration among them. 

\begin{figure*}
\centering
\includegraphics[width=.9\linewidth, angle=0]{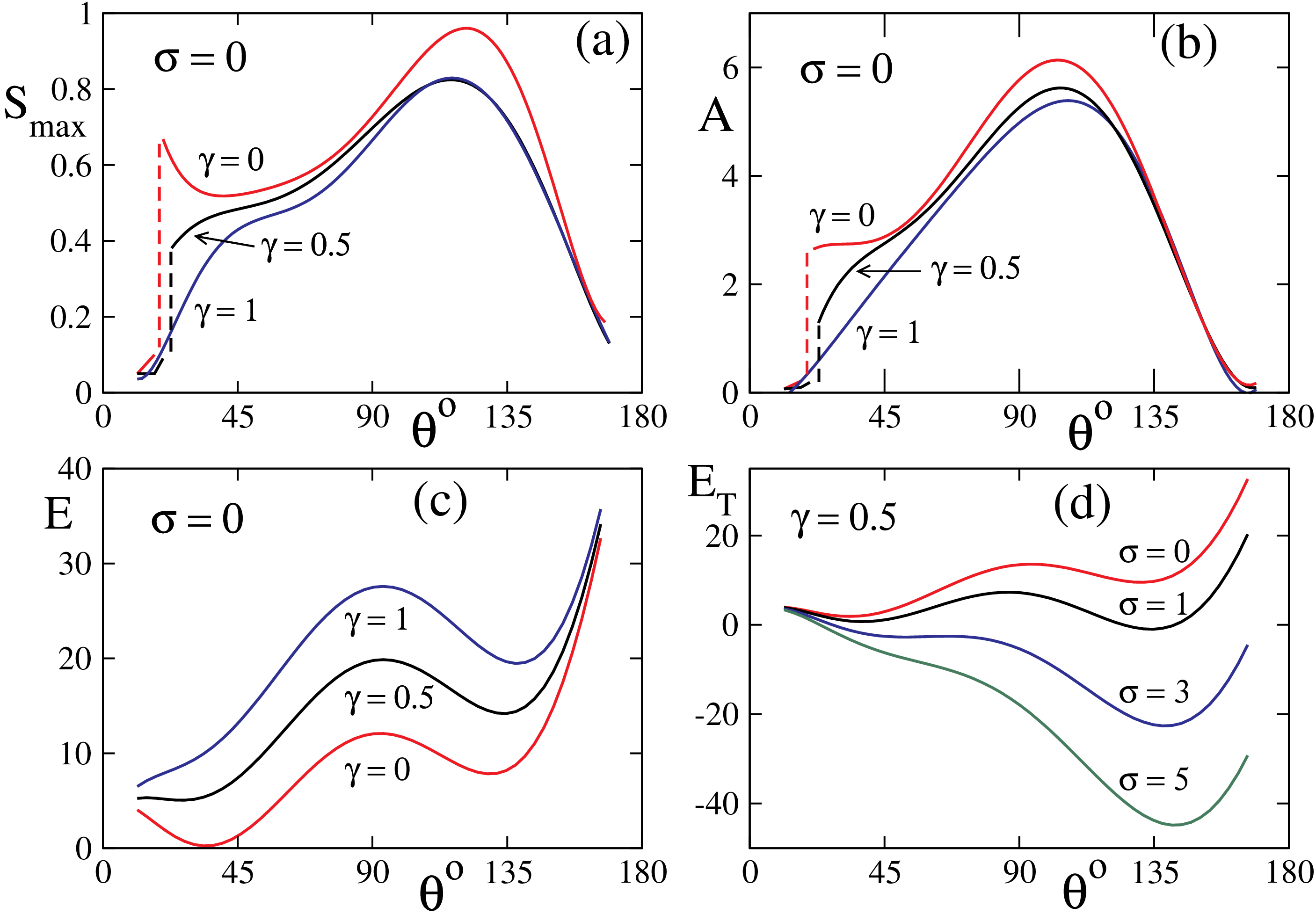}
\caption { a) Curve length $s_{max}$,  b) filament covered area $A$, and  c) energy $E$, shown as functions of $\theta_0$, at different values of $\gamma$, at $\sigma=0$, for $\sigma_1 <\mu$. d) Shows the corresponding total energy $E_T$ versus 
$\theta_0$, at $\gamma=0$, and for different values of $\sigma_1$.  
Membrane binding of the cargo can pull down the second minimum and make the bud stable. Dimensionless parameters : $k=10$, $\chi_0=0.75$, $\mu'=-0.5$. (a) and (b) exhibits a 1st order jump,
at $\gamma <1$.}
\label{fig.set3}
\end{figure*} 

\begin{figure*}
\includegraphics[width=.8\linewidth]{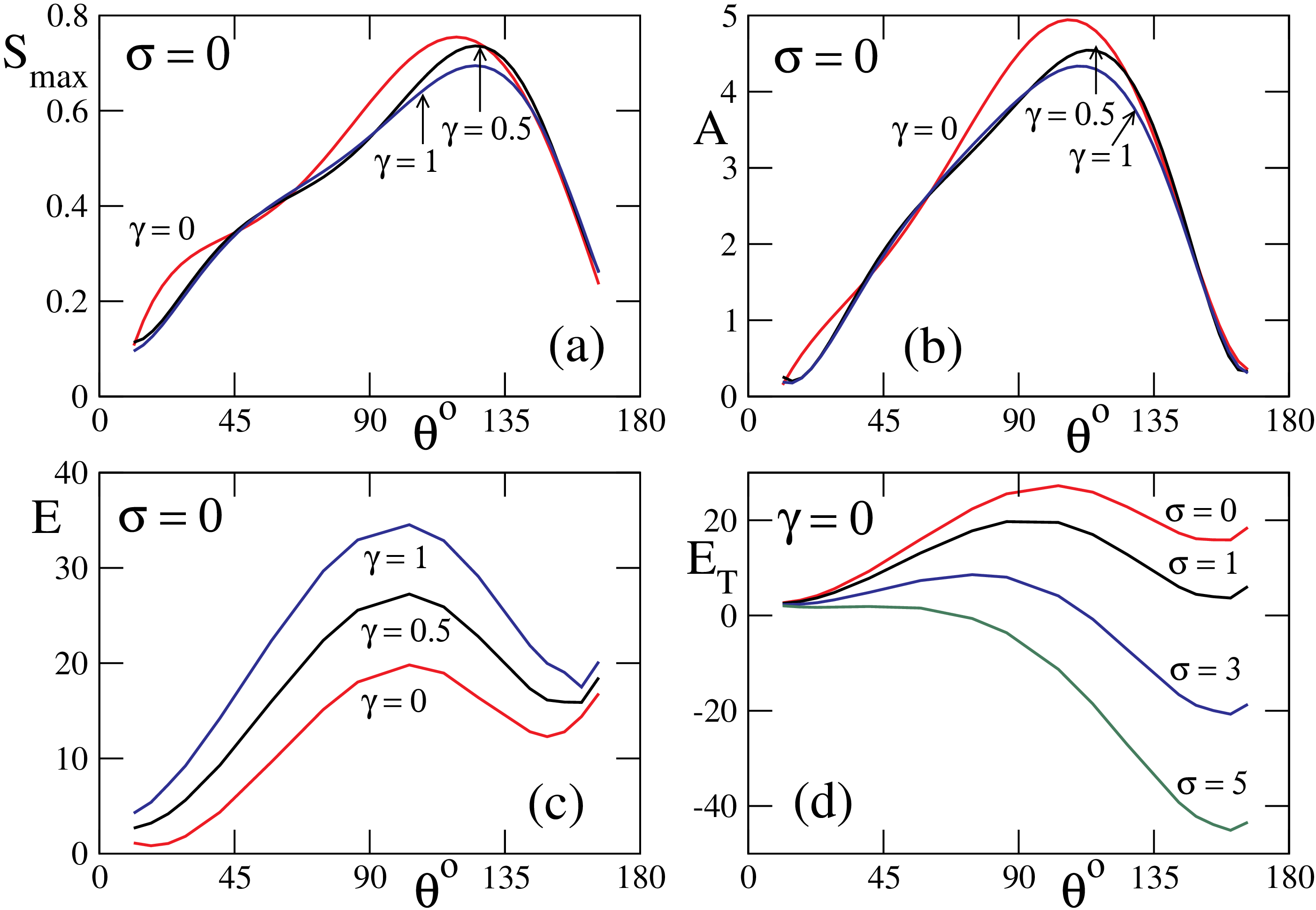}
 \caption{ a) Curve length $s_{max}$,  b) filament covered area $A$ and  c) energy $E$, plotted as a function of $\theta_0$, at different values of $\gamma$, at $\sigma=0$. d) Shows the total energy $E_T$ versus $\theta_0$, at $\gamma=0$ and for different values of $\sigma_1$.  But the second minimum has higher energy in the absence 
of adhesion energy due to the cargo and therefore cannot form a stable bud. Membrane binding of the cargo can pull down the second minimum and make the bud stable. Dimensionless parameters : $k=2.5$, $\chi_0=0.5$, $\mu' =0.5$}
 \label{fig.set1}
\end{figure*} 
\begin{figure}
\includegraphics[width=.9\linewidth]{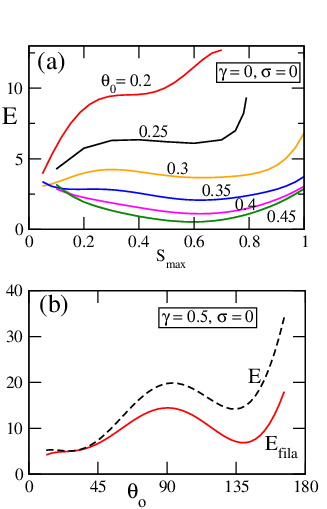}
 \caption{ Origin of 1st order jump : (a) Energy as a function of $s_{max}$ for different values of $\theta_0$ (in radian), for parameters corresponding to Fig.\ref{fig.set3}, at $\gamma=\sigma=0$. 
For a given  $\theta_0$, the equilibrium value of 
$s_{max}$ is the one that corresponds to the global minimum in energy. The figure shows a first order jump in the location of the global minimum, from $s_{max}< 0.1$ to $s_{max}\geq 0.6$, 
in a narrow range of $\theta_0$ around $0.3$ rad. 
b) Shows the contribution of the ESCRT-III filament (solid line) to the energy  $E(\sigma=0)$ (dashed line), as a function of $\theta_0$ seen in Fig.\ref{fig.set3}-c and Fig.\ref{fig.set1}-c. It implies that the double minima structure is due to the elastic energy of the filament.}
 \label{fig.jump}
\end{figure} 


\subsection{Phenomenological parameters }
We first estimate the various parameters for which we solve our model equations. 
The membrane bending rigidity $\kappa=20 \:k_BT$  is standard
for bilayer phospholipid membranes. 
Radius of the cargo is taken as $R=50 \:nm$, for most of the calculations. 

 Intrinsic radius of curvature $\chi_0$ for  ESCRT-III filaments have been estimated to be $25-30 \:nm$ by growing Snf7 (main component of ESCRT-III) filaments in vitro on supported membrane bilayers \cite{Lenz2015relaxation}.  These filaments nucleate as rings which 
later grow many concentric flat layers. Both electron microscopy (EM) 
and total internal reflection fluorescence (TIRF) microscopy 
measurements give similar results. Other studies (see references 
in \cite{Lenz2015relaxation}) also give similar estimates. 
In vivo measurements on cell membranes also show array of  
flat Snf7 concentric rings, in cells where Snf7 is over expressed.

Ref.\cite{Lenz2015relaxation} and references therein estimate
ESCRT-III persistence length $l_p$ in the range $250-800\:nm$. The 
bending modulus $k$ in our model, which has dimension of energy, 
can be extracted from $l_p$. In standard worm like chain (WLC) model filament bending energy $E=\frac{l_p.k_BT}{2}\int_0^L ds \: C^2(s)$, where the 
variable $s$ is the contour length and $C(s)$ is the local curvature 
along the contour of the filament. Introducing filament thickness 
$\Delta$ in the expression for energy we rewrite 
$E=\frac{l_p.k_BT}{2\Delta} \int dA \: C^2$, where $dA=\Delta ds$ is
the area element covered by a segment of the filament of length $ds$.
This yields $k=\frac{l_p.k_BT}{\Delta}$. Using $\Delta\approx 5nm$ \cite{hansenJBC08,Lenz2015relaxation} for width of the ESCRT-III single strand we estimate the dimensionless parameter $k/\kappa \approx 2.5-8$.
We note that ref\cite{lenz2009membrane} used $k/\kappa=2.5$. We
explore a range $k/\kappa=[2.5-10]$.

Adhesion strength of the cargo is not known precisely. We used the 
values of adhesion strength between the ESCRT-III filaments and 
the membrane, estimated in Ref.\cite{Lipowski2018domes}. They had 
estimated for both conical shaped as well as dome shaped necks. 
We use the one for cone shaped neck, see Fig. 6B in 
Ref \cite{Lipowski2018domes}, which in their notation $W\approx 0.05\:mN/m$.
In our dimensionless units $\sigma =WR^2/\kappa\approx 1.6$ and we 
explore a small range $[0-5]$ around this. Note that $W$ in  
Ref.\cite{Lipowski2018domes} is same as our $\mu$ which we had 
taken to be of order $\sigma_1=10^{-5}N/m$. Since the effective 
adhesion energy $\sigma=\sigma_0-\frac{\kappa}{2R^2}$ is further
reduced by $\frac{\kappa}{2R^2}$ which is of the same order, a 
low value $0.05\:mN/m$ is justified. 

We use membrane surface tension $\sigma_1=10^{-5}N/m$ from Ref.\cite{lenz2009membrane}. If we assume the membrane affinity
of ESCRT-III to be of similar order, then $-\mu' =\sigma_1-\mu$ will
also be of order $\sim 10^{-5}N/m$. Thus in dimensionless units  $\mu'R^2/\kappa\approx 0.3$. Ref.\cite{lenz2009membrane} has estimated $\sigma_1-\mu$ in a 
different way. They define a dimensionless effective surface 
tension $\Sigma= (\sigma_1-\mu)\frac{\chi_0^2}{\kappa} + \frac{k}{2\kappa}$
and considered different values of $\Sigma$. Taking $\Sigma=1$
we get $\mu'=1$. We explore a range $\mu'=[-0.5,0.5]$. In the 
regime where surface tension $\sigma_1$ is stronger $\mu'$ is 
negative and where membrane affinity of ESCRT-III $\mu$ is 
stronger $\mu'$ is positive. 

From Ref. \cite{lenz2009membrane}, we estimate $\gamma$ to be 0.6. We consider variation of $\gamma\in [0,1]$, to study its effect on the buckling process. Specifically, we use $\gamma=0,0.5$ and $1$. 

\section{Results}
We present below the solutions of the Euler-Lagrange's equations  obtained from this energy functional. We use experimentally relevant  values for the parameters $\kappa,k,\chi_0,\gamma$ and $R$, described above, and compute angular dependence of contour length $S_{max}$, filament covered area $A$ and the Energies  $E$ and $E_T$ (see Fig\ref{fig.set3},\ref{fig.set1}). We considered two broadly different scenarios : 
$\sigma_1 < \mu$ (Fig.\ref{fig.set3}) and $\sigma_1 > \mu$  (Fig.\ref{fig.set1}). Note that, $\sigma_1 < \mu$ can promote polymerization of ESCRT-III on the membrane and is presumably the expermentally relevant case, but direct measurement of $\mu$ is 
not available. Therefore we examine the $\sigma_1 > \mu$ case also.
It may appear that, $\sigma_1 < \mu$ would promote uncontrolled
polymerization at the base of the bud, however the consequent filament turns with progressively higher radius would cost higher energy due to the intrinsic curvature term and that would limit the outer radius of the coil. In Fig\ref{fig.set3}-d and \ref{fig.set1}-d the total energy ($E_T$) of the system is shown for different values of adhesion strength $\sigma$.
 

The common feature evident in the energy landscapes for both the cases, shown in  Fig\ref{fig.set3} and \ref{fig.set1}, is the existence of two minima. The minima at small and large $\theta_0$s' correspond to a nearly flat pre-bud and a fully formed bud, respectively. Accordingly, the shape of the ESCRT-III spiral is nearly flat at the first minima and conical at the second minima. Plots of 
$E$ in Fig\ref{fig.set3},\ref{fig.set1}-c reveals that without the contribution from the favorable adhesion energy of the cargo ($E_c$) the bud (the second minima) is metastable since it has higher energy compared to the 1st minima corresponding to the pre-bud state. Also since the 1st minima occurs at a finite $\theta_0$, it is attained via the buckling of the flat membrane at $\theta_0=0$. But the energy of the filament and the membrane associated with it cannot drive this buckling, except for $\gamma\simeq 0$ in the $\sigma_1<\mu$ case (see Fig.\ref{fig.set3}-c). Therefore favorable contribution from adhesion energy is crucial even for this initial buckling transition.
Furthermore, comparison between (c) and (d) in both Fig\ref{fig.set3},\ref{fig.set1} shows that there is a barrier between the nominally buckled pre-bud state and the metastable bud, which can be overcome only by a strong adhesion energy. 

Formation of the pre-bud at small finite $\theta_0$ shows qualitative difference between the $\sigma_1<\mu$ and $\sigma_1>\mu$ cases. For the $\sigma_1<\mu$, the pre-bud is a well defined minima of $E_T$ while for $\sigma_1>\mu$ it is not. Although in both cases there is no jump in the energy $E$ or $E_T$ around the pre-bud state, for $\sigma_1<\mu$ there is a discontinuous jump in both $s_{max}$ and $A$, at $\gamma <1$. But physically, this region of small $\theta_0$ corresponds to the nucleation of the ESCRT-III spiral, which could 
a more complicated activated process, not captured by our model. Nevertheless, we analysed the origin of this jump in $s_{max}$ in 
Fig.\ref{fig.jump}-a. The lowest minima at a given $\theta_0$ correspond to the equlibrium value of $s_{max}$ (shown in Fig.\ref{fig.set3}). In Fig.\ref{fig.jump}-a, we see that, as $\theta_0$ is increased the boundary minima at $\theta_0$ loose out to the 
developing minima at finite $\theta_0$ in the range $[0.3-0.35]$,
causing the system to jump to the new global minima, as it occurs 
for a 1st order transitition. For example, a mean field theory 
with a cubic order parameter term gives abrupt jump in magnetization,
from zero to a nonzero value, although the free energy changes smoothly at the transition.

The double minima structure in the elastic energy of the filament (Fig.\ref{fig.jump}-b) is due to its intrinsic curvature. It can 
be understood by considering contribution 
coming from the top most part of the filament which has radius $X(0)=R\sin\theta_0$. With $\chi_0=25nm$ and $R=50nm$ the 
elstic energy contribution is proportional $(\frac{1}{\sin\theta} -2)^2$
which has a double well structure with a maxima at $\theta=\pi/2$. 
This is because as $\theta_0$ increases from $0$ to $\pi$, initailly $X(0)$ is smaller than $\chi_0$, then it increases and goes past  $\chi_0$ and then again it decreases and falls below 
$\chi_0$. The energy is high whenever $|X(0)-\chi_0|$ is high.
The rest of the conical shaped filament also behave similarly.

In Fig.\ref{fig.jump}-b note that $E-E_c$, the elastic energy contribution of the membrane, increases slowly as its 
curvature increases with $\theta_0$ (as the bud develops).
Also note that the area associated with the filament first 
increases and then later decreases (see subfigure (a) in both 
Fig.\ref{fig.set3},\ref{fig.set1}). This indicates that polymerization in the beginning and depolymerization near 
the end as we did not constrain the  length of the filament 
in our model.
We now discus how the different features of the energy landscapes shown in Fig.\ref{fig.set3},\ref{fig.set1} changes as we change 
the various parameters of the model. 

\begin{enumerate}

\item \textbf{Line tension $\gamma$ :}  
Fig.\ref{fig.set3}-a,b, show that, for the $\sigma_1<\mu$ case, the jump associated with the contour length ($s_{max}$) and the filament-covered membrane area ($A$) gradually disappear as $\gamma$ is increased. However, for the  $\sigma_1>\mu$ case, $s_{max}$ and $A$ do not change much, but the energy $E$ changes significantly with $\gamma$. In particular the barrier between the pre-bud and bud increases with $\gamma$ and therefore stronger adhesion ($\sigma$) is needed to cross this barrier. Althugh the position of the maxima 
(in $\theta_0$) does not change much with $\gamma$, the position 
of the second minima shifts right with increase in $\gamma$.. As mentioned before $\gamma$ does not explicitly enter the Euler-Largarne equations (Eq.\ref{eq.EL}) but it influences the energy $E$ (see Eq.\ref{eq.E}) and thereby influences the equilibrium value of $s_{max}$ for a given $\theta_0$.

\item \textbf{Filament stiffness $k$ :} Comparison between Fig.\ref{fig.set3} (with $k=10$) and Fig.\ref{fig.set1} (with $k=2.5$) shows that, as $k$ decreases, the second minima shifts to 
larger $\theta_0$ i.e., closer to scission ($\theta_0=180^o$)
(figure not shown). At the same time the energy barrier for 
scission of the bud also goes down. But note that our model does 
not address scisson which could be due to a nonequilibrium process. 

\item \textbf{Intrinsic filament curvature $\chi_0$ :} Decrease
in $\chi_0$ reduces the jump associated with curve length and area,
for the $\sigma_1<\mu$ case (figure not shown). Also the second minima shifts towards higher $\theta_0$ helping the scission process.

\item \textbf{Cargo size $R$ :} We checked that at larger cargo
size (eg., $R=100nm$) the jumps in $s_{max}$ and $A$ disappear,
and the  second minima shifts rightward (figure not shown). This 
is because the process gets completely dominated by the adhesion energy of the cargo. 
\end{enumerate}

\section{Discussion}

The common theme which emerges is that, a) the transition of ESCRT-III from flat spiral to conical spiral is not driven by the intrinsic cirvature of ESCRT-III filaments (Fig.\ref{fig.jump}-b). b) The evolution of mildly deformed pre-bud
like structure to a bud structure requires drive from adhesion energy. c) This pre-bud structure and the final complete bud structure constitute two energy minima separated by a barrier whose magnitude depends on the adhesion energy and at strong adhesion the barrier can even vanish clearing way for a smooth passage to complete bud like structure. However even in the presence of a finite barrier energy, non-equilibrium processes \cite{Saric2019changes} can help in the barrier crossing. We did not use any 
conformational transition in the orientation of the molecular bonds in the filaments, which can generate tilt in the turns of the filament promoting a helix, as proposed in Ref \cite{Saric2019changes}. 

\begin{figure}
   \centering
    \includegraphics[width=0.85\linewidth, angle=0]{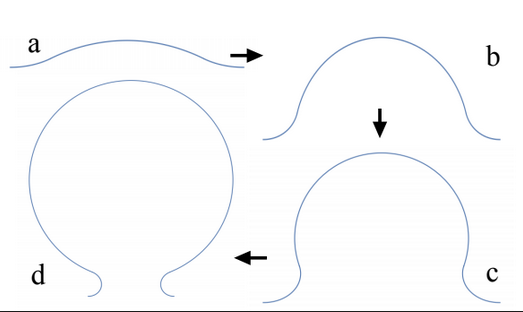}
    \caption{Evolution of surface with increasing initial surface angle for parameter set:  $k=10$, $L=0.75$ and $\mu' = - 0.5$}
    \label{fig:revolution}
\end{figure}

Note that the adhesion of the cargo to the cell membrane breaks the up-down symmetry of the membrane deformation and makes the filament buckle in the 
direction away from the cell body (convex) instead of towards the interior 
of the cell body (concave). However, without the adhesion, in our model the membrane is absolutely symmetric with respect to  budding out or budding in.
In case of inward budding the ESCRT-III filament would wrap on the neck of
the bud on its outer surface which may not be possible in reality as ESCRT-III is always known to cut membrane necks from inside. But in our model we do not distinguish between filament wrapping the membrane from outside or inside.

In our model the length of the filament is not fixed.
The filament has an affinity for the membrane and it is found to polymerize first and then depolymerize near the completion of the process which is consistent with observations. 

A relevant question has been what promotes the filament to become a conical spiral instead of a helical spiral. 
Our model shows that the shape is neither a uniform cone nor a cylinder. It is worth noting that, i) the filament starts out as a flat spiral 
\cite{hansenJBC08}, ii) a helical spiral requires sharp membrane bending at the base, and iii) as adhesion of the cargo increases the radius 
of the neck reduces below $\chi_0$ promoting a conical structure.


It is not evident from experiments whether the filament grows as a flat spiral first on the membrane and then it transits to a helical shape as the bud starts to grow or does it polymerize
only when the budding starts, driven by cargo-membrane adhesion. Question arises, in the first case if the membrane surface is already covered by a flat spiral then how does the cargo adhere to the membrane. Our model assumes the
second possibility that the polymerization starts with the budding.

Note that our model does not answer how membrane scission takes
place at the end of budding. After the second minima the total energy $E$ starts to rise sharply (see Fig.\ref{fig.set3}) due to the bending energy costs of both the membrane and the filament, making the minima stable at an angle $\theta_0<\pi$. This asks for a separate mechanism for scission of the bud.

Also, we have implicitly assumed that the filament-spiral ends where the membrane touches the flat surface at $s_{max}$, which means the filament is not preferred on flat surface. This is certainly true for 
$\sigma_1<\mu$, which is experimentally relevant.



\bibliographystyle{unsrt}
\bibliography{buckling}
\end{document}